\documentclass[aps,pre,floatfix,showpacs,twocolumn, preprintnumbersm,nofootinbib]{revtex4-1} %

\usepackage{graphicx}
\usepackage{amssymb}
\usepackage{amsmath}
\usepackage{mathtools}
\usepackage[dvipsnames,usenames]{xcolor}
\usepackage{multirow}
\usepackage{bm}
\usepackage[hidelinks]{hyperref}
\usepackage[dvipsnames]{xcolor}
\usepackage{xspace}
\usepackage{multirow}

\newcommand{\figt}{F{\footnotesize IG}\xspace}
\newcommand{\figts}{F{\footnotesize IG}s\xspace}

\newcommand{\lcdm}{$\Lambda$CDM\xspace}
\newcommand{\lya}{Ly$\alpha$\xspace}
\newcommand{\jerald}{\texttt{JERALD}\xspace}

\newcommand{\Fref}{F_\text{ref}\xspace}
\newcommand{\dstar}{\delta_\text{star}\xspace}
\newcommand{\dout}{\delta_\text{out}\xspace}
\newcommand{\rout}{\rho_\text{out}\xspace}

\newcommand{\khSM}{k^h_{\text{sm}}\xspace}
\newcommand{\klSM}{k^l_\text{sm}\xspace}
\newcommand{\nSM}{\nu_\text{sm}\xspace}
\newcommand{\khDM}{k^h_\text{dm}\xspace}
\newcommand{\klDM}{k^l_\text{dm}\xspace}
\newcommand{\nDM}{\nu_\text{dm}\xspace}
\newcommand{\khI}{k^h_\text{I}\xspace}
\newcommand{\klI}{k^l_\text{I}\xspace}
\newcommand{\nI}{\nu_\text{I}\xspace}

\begin{document}

\title{\jerald: high-fidelity dark matter, stellar mass and neutral hydrogen maps from fast N-body simulations}
\author{ {Mauro} Rigo$^{\, {\rm a} }$, {Roberto} Trotta$^{\,\rm {a, b, c, d}}$, and {Matteo} Viel$^{\, {\rm a, b, c, e} }$ 
}

\affiliation{
$^{\rm a}$\mbox{Theoretical and Scientific Data Science, SISSA, Via Bonomea 265, 34136 Trieste, Italy}\\
$^{\rm b}$INFN -- National Institute for Nuclear Physics, Via Valerio 2, 34127 Trieste, Italy\\
$^{\rm c}$ICSC - Centro Nazionale di Ricerca in High Performance Computing, Big Data e Quantum Computing,
Via Magnanelli 2, Bologna, Italy \\
 $^{\rm d}$Astrophysics Group, Physics Department, Blackett Lab, Imperial College London, Prince Consort Road, London SW7 2AZ, UK \\
 $^{\rm e}$INAF -- Osservatorio Astronomico di Trieste, Via G. B. Tiepolo 11, I-34143 Trieste, Italy\\
}

\date{\today}

\begin{abstract}
We present a new code and approach, \jerald---JAX Enhanced Resolution Approximate Lagrangian Dynamics---, that improves on and extends the Lagrangian Deep Learning method of Dai \& Seljak (2021), producing high-resolution dark matter, stellar mass and neutral hydrogen maps from lower-resolution approximate $N$-body simulations. The model is trained using the Sherwood-Relics simulation suite (for a fixed cosmology), specifically designed for the intergalactic medium and the neutral hydrogen distribution in the cosmic web. The output is tested in the redshift range from $z=5$ to $z=0$ and the generalization properties of the learned mapping is demonstrated. \jerald produces maps with dark matter, stellar and neutral hydrogen power spectra in excellent agreement with full-hydrodynamic simulations with $8\times$ higher resolution, at large and intermediate scales; in particular, \jerald's neutral hydrogen power spectra agree with their higher-resolution full-hydrodynamic counterparts within 90\% up to $k\simeq1\,h$Mpc$^{-1}$ and within 70\% up to $k\simeq10\,h$Mpc$^{-1}$. \jerald provides a fast, accurate and physically motivated approach that we plan to embed in a statistical inference pipeline, such as Simulation-Based Inference, to constrain dark matter properties from large- to intermediate-scale structure observables.
\end{abstract}

\maketitle

\section{Introduction}

Cosmological simulations are essential tools for studying the universe, as modeling physical observables is necessary to extract the maximum amount of information from current and future sky surveys such as DESI \cite{desi}, Euclid \cite{euclid}, SKA \cite{ska}, and LSST \cite{lsst}. These surveys will probe vast volumes of the sky over a broad frequency spectrum, gathering data for a great variety of phenomena and observables such as galaxy clustering, weak lensing, supernovae type Ia, 21-cm radiation, and the \lya forest with unprecedented resolution. As an example, Euclid will cover more than a third of the sky in its 6-years mission, with angular resolutions up to few tenths of arcseconds, while LSST, in 10 years, will study over 40\% of the sky with similar resolution. Such impressive numbers will allow to investigate the universe over very small scales, which offer a multitude of avenues for scientific research. For example, small-scale (putative) problems yet to be solved such as the core/cusp problem \cite{core-cusp2, core-cusp3} and the missing satellite problem \cite{missing-satellites1, missing-satellites2} challenge our current understanding of the process of structure formation, and while solutions may exist within the \lcdm paradigm, these inconsistencies constitute strong drivers for research beyond the standard cosmological model \cite{smallscalesprob}.

Yet, the exciting opportunities that these surveys offer pose a major problem: on scales $\lesssim 1\,h^{-1}$Mpc and particularly at late times, perturbations become highly nonlinear, thus requiring high-resolution simulations to be modeled accurately. The most advanced simulations, called full-hydrodynamical simulations owing to their treatment of baryons as fluids, are capable of tracking the evolution of several billions of dark matter (DM) and baryonic particles for several billions of years, modeling many of the physical phenomena that govern their dynamics on a wide range of scales \cite{TNG, camels, Sherwood,sherwood-relics}; nonetheless, they are computationally extremely expensive, often requiring millions of CPU hours \cite{TNGtime, Sherwood}. One of the major culprits for the enormous computational cost of numerical simulations is baryonic matter itself, since baryons are characterized by a vast range of physical processes that need to be implemented in order to accurately describe their interactions. On the contrary, DM is relatively inexpensive to simulate as it only interacts through gravity, which can either be solved fully or even approximately with a computational complexity of $\mathcal{O}(N \log N)$ for $N$ particles \cite{fastpm, cola}. Therefore, since gravity is the dominant component over large scales, where baryons trace the evolution of DM, one possible approach for mitigating the great computational cost of full-hydrodynamical simulations consists in running either full or approximate DM-only ($N$-body) simulations, and then populating them with baryons or extracting baryonic properties in a post-processing step. 

Many different approaches have been proposed in this direction: semi-analytical methods \cite{2002ApJ...575..587B, baugh06, universemachine, dm-to-galaxy1, dm-to-galaxy2, sage}, including abundance matching techniques \cite{sham1, sham2, sham3}, are used to reproduce galaxy and halo properties, while in recent years machine learning (ML)-based methods have flourished, including e.g.\ support vector machines, $k-$nearest neighbor algorithms or tree-based methods \cite{ml-mock-galaxy-catalogs, tree-based, tree-based2, 10.1093/mnras/sty1169}. Other authors treat the problem of painting baryons on DM-only simulations as an image-to-image or an image generation task, developing pixel-based methods to map DM distributions to different baryonic observables using convolutional neural networks (CNNs) \cite{zhang2019dark, Villaescusa-Navarro_2023, 10.1093/mnras/stad2596, unet-wgan}, sometimes in conjunction with variational diffusion models (VDMs) \cite{cnn-vdm}, or using variational auto-encoders (VAEs) and generative adversarial networks (GANs) \cite{10.1093/mnrasl/slz075} as well as score-based models \cite{score-based}. 

In this work we follow another approach, which is well-suited for problems with relatively simple large-scale physics but complex small-scale effects that can be approximately modeled using an effective theory framework, and it consists in rewriting the Lagrangian describing the system in a way that explicitly satisfies the symmetries of the problem, but parametrized by a set of free parameters meant to capture non-perturbative small-scale effects. Contrarily to pixel-based approaches---which treat DM and baryons distributions as images and therefore do not exploit intrinsic physical symmetries---this ``Lagrangian approach'' works at the level of individual particles or fluid elements of simulations by modeling their displacement field through an effective potential. 

In the context of cosmological simulations, this scheme has a straightforward application in improving approximate $N$-body simulations, where the Lagrangian framework is used to compensate for the loss of power of the DM distribution at small scales caused by the coarse-graining in time and space in these types of simulations \cite{PGD}. By writing the effective potential used to displace the particles in a suitable form, this technique preserves the translational and rotational symmetries, while simply adjusting the location of the particles explicitly satisfies mass conservation. In an approach they called \textit{Lagrangian Deep Learning}\footnote{This name was motivated by the presence of multiple displacement layer in the original implementation. However, `deep learning' as usually understood in a machine learning sense involves a large number of free learnable parameters, which is not the case here; this is why we prefer to refer to our implementation in terms of `approximate Lagrangian dynamics'.} (LDL), Dai and Seljak \cite{LDL} propose to extend this idea to baryons and their properties. Since baryons trace the evolution of dark matter on large scales and the two distributions differ mostly on small scales, in the Lagrangian framework the distribution of dark matter can be thought of as a large-scale distribution of baryons, and the small-scale baryonic physics can instead be directly modeled at the desired redshift by displacing the DM particles, reinterpreted as baryons, such that their final density matches the one of a specific baryon type from some reference full-hydrodynamical simulation. Additionally, the authors extended the model to predict other baryonic observables such as thermal and kinetic Sunyaev-Zeldovich signals and X-ray emission.   

Building on this idea, in this paper we present a new code and an improved approach, under the name of `JAX Enhanced-Resolution Approximate Lagrangian Dynamics' or \jerald, which increases the computational efficiency of LDL and demonstrates the generalization properties of the method to other simulation suites, to higher redshift and to the modeling of neutral hydrogen (HI). 

This paper is organized as follows. In Section \ref{sec:model} we introduce \jerald and the improvements it brings over LDL; we also explain how neutral hydrogen is modeled within our framework. In Section \ref{sec:code} we give details about the implementation and the performance of the code. In Section \ref{sec:results} we present our validation results, as well as generalization to different initial conditions. Concluding remarks and future perspectives are given in Section \ref{sec:conclusions}.

\section{The model}
\label{sec:model}

In this section, we describe the \jerald model by detailing our new implementation of the original idea by Dai and Seljak \cite{LDL}, and describing the pipeline of the approach. In Sec.\,\ref{sec:LDL}, we first explain how the DM distributions from approximate $N$-body simulations are improved, also to account for baryonic feedback, and how stellar mass distributions are learned, while the case of neutral hydrogen is treated separately in Sec.\,\ref{sec:nHI}, as it poses specific challenges that require taylored solutions. 

\subsection{Lagrangian approach and \jerald}
\label{sec:LDL}

\begin{figure}
    \centering
    \includegraphics[width=\linewidth]{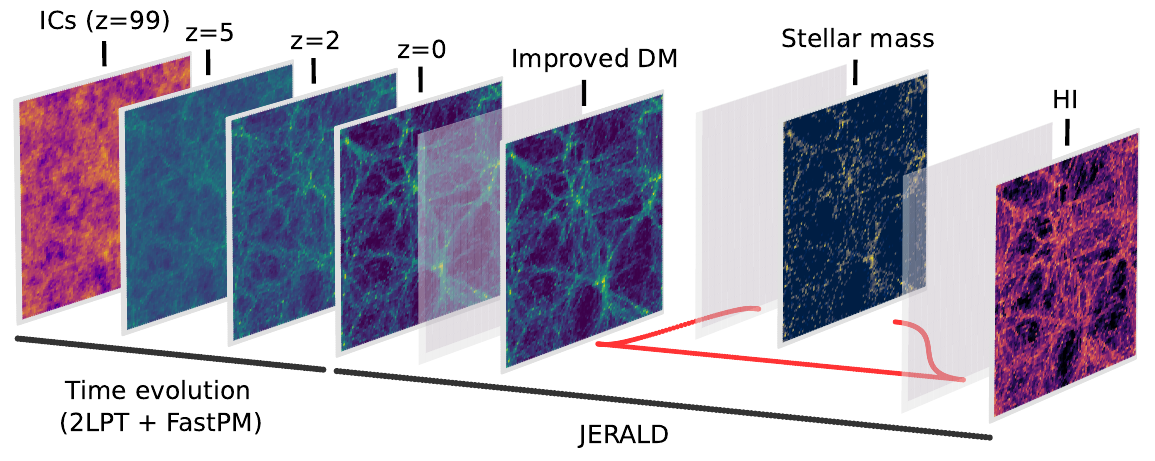}
    \caption{Schematic of the \jerald pipeline: the FastPM approximate $N$-body solver starts from initial conditions at high redshift ($z=99$), evolves DM particle positions first via 2LPT (up to $z=9$) and then, though a small number of leapfrog integration steps, produces a final array of DM particle positions at $z=5$, $z=2$ and $z=0$, the cases analyzed in this work. Starting from these positions---here, we illustrate the case of $z=0$---\jerald improves the DM distribution by optimizing the free parameters of the displacement layers on the DM target from a full-hydrodynamic simulation. Then, \jerald applies additional displacement layers to the improved DM positions, and a final transformation to produce a stellar mass distribution, with parameters optimized on the stellar mass target. Finally, starting again from the improved DM particles positions and using the pre-trained stellar mass map, \jerald produces a HI map by displacing the particles via Eq.~\eqref{eq:displacementHI} and applying the transformation in Eq.~\eqref{eq:reluHI}. Gray layers represent additional intermediate displacement layers.}
    \label{fig:pipeline}
\end{figure}

Like LDL, \jerald takes as input DM particles positions, moves them via one or more layers of Lagrangian displacement (in this paper, we mostly use 2), and finally produces as output a density field via a mass-assignment scheme (Cloud-In-Cell or CIC), optionally applying a non-linear transformation to the output. Given a particle $i$ with position $\bm r_i = (r_i^x,\,r_i^y,\,r_i^z)$ in comoving coordinates, a Lagrangian displacement layer acts as
\begin{equation}
    \bm r_i \to \bm r_i + \bm S(\bm r_i) \ ,
\end{equation}
where the displacement field $\bm S(\bm r)$ is modeled as the gradient of an effective potential, the source of which, similarly to gravity, is a simple power law of the normalized density:
\begin{equation}
\label{eq:displacement}
    \bm S = \alpha \bm\nabla \big(\hat Gf(\delta)\big)\quad,\qquad f(\delta)=(1+\delta)^\gamma\ ,
\end{equation}
where we omitted the dependence on $\bm r$ for ease of notation and
\begin{equation}
    1+\delta(\bm r)=\rho(\bm r)/\sum_i \rho(\bm r_i)
\end{equation}
is the normalized density, computed assuming the same mass for each particle (which cancels out when normalizing by the average density).
In Eq.~\eqref{eq:displacement}, $\alpha$ and $\gamma$ are free (learnable) parameters and $\hat G$ is the Green's operator, the action of which can be explicitly written as
\begin{equation}
\label{eq:convol}
    \hat Gf(\delta)(\bm r) = \int G(\bm r, \bm r')f(\delta(\bm r'))d\bm r' \ ,
\end{equation}
with $G(\bm r, \bm r')=G(\bm r - \bm r')$ because of translational symmetry. Being a convolution, Eq.~\eqref{eq:convol} has a straightforward expression in Fourier space as
\begin{equation}
    \mathcal{F}\big[\hat Gf(\delta)\big](\bm k)=\Tilde G(\bm k)\, \Tilde f(\delta)(\bm k)\ ,
\end{equation}
with $\Tilde G(\bm k)=\Tilde G(k)$ due to rotational symmetry.

The explicit form of $\Tilde G(k)$ depends on the target quantity: when modeling stellar mass, the Green's operator is parameterized, as in LDL, in the following way:
\begin{equation}
\label{eq:greenSM}
    \Tilde G_\text{sm} (k)=\exp\left[-\left(\frac{\khSM}{k}\right)^2\right]\exp\left[-\left(\frac{k}{\klSM}\right)^2\right]k^{\nSM} \ ,
\end{equation}
where the first factor suppresses large-scale growth for $k < \khSM$, since the physics to model through an effective description happens at small scales, while the second term mimics force softening. The final power law term is motivated by the fact that fixing $\alpha=4\pi G\Bar{\rho}$ and $\gamma=1$ in Eq.~\eqref{eq:displacement}, setting $\khSM=0$, $\klSM=\infty$ and $\nSM=-2$ in Eq.~\eqref{eq:greenSM} retrieves the gravitational potential. Here, $\khSM$, $\klSM$ and $\nSM$ are again free (learnable) parameters. 

In the case of DM, we find that a slightly different expression for the Green's function works best:
\begin{equation}
\label{eq:greenDM}
    \tilde G_\text{dm}(k)=\exp\left[-\frac{\khDM}{k}\right]\exp\left[-\frac{k}{\klDM}\right]k^{\nDM} \ .
\end{equation}

In practice, each displacement layer is implemented as follows. Given particle positions $\bm r$, a density field is produced via CIC and raised to some power (learned and different between displacement layers), as indicated in Eq.~\eqref{eq:displacement}. The potential in Fourier space is then computed as the product of the Discrete Fourier Transform (DFT) of this field and the Green's function parameterized based on the quantity to model---i.e. as in Eq.~\eqref{eq:greenSM} for stellar mass, or as in \eqref{eq:greenDM} for DM. The gradient is calculated using the finite difference method; since, in real space, a finite difference approximation of a derivative can be written as a convolution, this simply translates into a multiplication by a filter in Fourier space. Finally, the gradient of the potential is transformed back into real space and interpolated from the grid points of the CIC mesh to the particle positions $\bm r$, using an interpolation scheme analogous to CIC.

After the displacement layers, a density map $\rout$ is produced again using CIC. The final transformation applied to $\rout$ depends on the target map type and is designed to capture the physical processes that cannot be modeled simply via matter transport, such as star formation for a stellar mass target and ionization for neutral hydrogen. 

For stellar mass, the transformation has the form:
\begin{equation}
\label{eq:reluSM}
    F_\text{sm}(\bm r) = \text{ReLU}(w_\text{sm}(1+\dout(\bm r))^{\mu_\text{sm}} - b_\text{sm})\ ,
\end{equation}
where $w_\text{sm},\,\mu_\text{sm}$ and $b_\text{sm}$ are free (learnable) parameters and $\dout$ is the overdensity associated to the density $\rout$. This transformation is meant to select regions of high density where star formation occurs. 

In the case of DM, no final transformation is applied, implying
\begin{equation}
\label{eq:Fdm}
    F_\text{dm}(\bm r)=1+\dout(\bm r) \ .
\end{equation}

The free parameters of the model are optimized by matching the output field $F(\bm r)$ to a distribution of the physical quantity to be modeled $F_\text{ref}(\bm r)$, obtained from a high-resolution full-hydrodynamical reference simulation. This is achieved by minimizing an L1 or L2 loss:
\begin{equation}
\label{eq:loss}
    \mathcal{L}=\frac{1}{N_\text{grid}}\left\|\Hat O_s F(\bm r) - \Hat O_s \Fref(\bm r)\right\|^p\ ,
\end{equation}
where $\|\cdot\|^p$ is the $p$-th norm ($p=1,\,2$), $N_\text{grid}$ is the number of grid points (or voxels) in the maps $F$, $\Fref$ and $\Hat O_s$ is an optional smoothing operator (used for the stellar mass maps and HI at $z=5$ only), defined in Fourier space as
\begin{equation}
\label{eq:loss2}
    \Tilde O_s(k) = 1+\left(\frac{k}{1h\text{Mpc}^{-1}}\right)^{-n}\ ,
\end{equation}
where $n$ is a hyperparameter that determines the relative weight between the large-scale and small-scale modes, introduced to prevent the model from focusing predominantly on the small scales. Similarly to the original implementation, in the case of DM, we use an L2 loss and no smoothing, while for stellar mass we use an L1 loss and a different $n$ for different redshifts, as we shall explain below.

\subsection{Predicting neutral hydrogen}
\label{sec:nHI}

Going beyond \cite{LDL}, we now extend the model to predict HI maps. In this case, the key challenge is represented by galactic feedback, a complex process whose impact on the HI distribution propagates from small scales, where it originates, to large scales. The role of feedback on the HI distribution is similar to that on the overall gas distribution: at small/intermediate scales, galactic winds are likely to push gas outside galaxies in the low-density intergalactic medium (IGM), either with thermal or mechanical feedback, while at small scales baryons can condense and determine an increase of power compared to a simulation in which baryons are absent (see e.g. \cite{vandaalen20}). HI is expected to follow, at least to some extent, the gas distribution, providing an overall suppression of power at small/intermediate scales and increase of power at very small scales \cite{vandaalen11,schneider19,parimbelli19}; however, differently from other baryonic components, the HI distribution is also sensitive to the local gas temperature, UV background and radiative transfer effects from astrophysical sources that ionize the surrounding medium, including patchy reionization \cite{sherwood-relics}. 

Furthermore, the dynamical range of HI is much greater compared to that of, e.g., stellar mass, and it can stretch over several orders of magnitude, from neutral fractions of $10^{-7}$ in filaments up to fully neutral gas inside galaxies in self-shielded regions at low redshift \cite{villa18}: for this reason, devising an approximate model that can reproduce the HI distribution in its entire dynamical range is a very challenging task. Nonetheless, different parts of this range can be probed through different observables: in particular, the high-density regions, found within halos (i.e., at $z=0$, with number densities above $\sim10^{-10}$cm$^{-3}$ and up to $\sim10^{-5}$cm$^{-3}$), are observable via 21cm intensity mapping, while the low to intermediate densities, found in the IGM (at $z=0$, with HI number density reaching down to $\sim10^{-17}$cm$^{-3}$ at $z=0$), are relevant for the study of the \lya forest.

In this work, we focus on the IGM; clearly, using the loss in Eq.~\eqref{eq:loss} to train \jerald on a reference simulation that models the full dynamical range of HI without any pre-processing would result in the model focusing solely on the high-density range, which would dominate the loss. This problem can be circumvented in various ways: for example, by clipping the value of the HI density above a given threshold in the reference simulation, thus allowing the cosmic web to stand out and be learned. Here, this step is not necessary, as the reference simulations utilized do not implement self-shielding \cite{Sherwood}: this results in the HI inside halos being greatly underestimated, which is beneficial for our purpose as it reduces the dynamical range. More generally, if the main focus is the IGM, the features of the distribution one has to model depend on the prescription used to mask out the high-density regions: the approach described below is therefore well-suited for reference simulations without self-shielding, while, for example, clipped maps would require a different treatment---we plan to explore this in the future.

In practice, to model neutral hydrogen, we modify the original LDL approach in two ways. Firstly, we introduce a new term in the effective potential:
\begin{equation}
\label{eq:displacementHI}
    \bm S = \alpha^{(1)}_\text{HI} \bm\nabla \big(\hat G^{(1)}_\text{HI}f^{(1)}_\text{HI}(\delta)\big) + \alpha^{(2)}_\text{HI} \bm\nabla \big(\hat G^{(2)}_\text{HI}f^{(2)}_\text{HI}(\dstar)\big)\ ,
\end{equation}
where $\alpha^{(1)}_\text{HI}$ and $\alpha^{(2)}_\text{HI}$ are free parameters, both Green's operators $\hat G^{(1)}_\text{HI}$ and $\hat G^{(2)}_\text{HI}$ are defined as in Eq.~\eqref{eq:greenSM} and have separate free parameters, $f^{(1)}_\text{HI}$ and $f^{(2)}_\text{HI}$ are power laws with separate free exponents as in Eq.~\eqref{eq:displacement}, $\delta$ is once again the input DM overdensity while $\dstar$ is the stellar mass overdensity associated to the distribution $\delta$. The purpose of the second term in Eq.~\eqref{eq:displacementHI} is to approximately model the effect of galactic winds. In order to predict HI using solely the input DM particle positions, for $\dstar$ we use the prediction obtained from \jerald as described in the previous section and pre-trained separately, i.e. $1+\dstar=F_\text{sm}$ from Eq.\,\eqref{eq:reluSM}. 

Secondly, we replace the final transformation of Eq.~\eqref{eq:reluSM} with the expression:
\begin{equation}
\label{eq:reluHI}
    F_\text{HI}(\bm r) = \text{ReLU}(w_\text{HI}(1+\dout(\bm r))^{\mu_\text{HI}} - D(\bm r) - b_\text{HI})\ ,
\end{equation}
where $D(\bm r)$ is a depletion term given by
\begin{equation}
    D = \text{ReLU}(\beta^{(1)}\hat O_I\, (1+\dstar)+\eta) + \beta^{(2)} f^{(3)}(\dstar)\ ,
    \label{eq:depletion}
\end{equation}
where $\beta^{(1)}$, $\beta^{(2)}$ and $\eta$ are free parameters, $f^{(3)}$ is once again a power law as in Eq.~\eqref{eq:displacement} with a different free exponent and $\hat O_I$ is a filter defined in Fourier space as
\begin{equation}
\label{eq:dfilter}
    \Tilde O_I(k)=\exp\left[-\frac{\khI}{k}\right]\exp\left[-\frac{k}{\klI}\right]k^{\nI} + \xi \ ,
\end{equation}
with $\khI$, $\klI$, $\nI$ and $\xi$ learnable parameters. Heuristically, the first term in Eq.~\eqref{eq:depletion}, which affects the HI distribution both where stars are present and in their surroundings, is meant to capture the depletion of HI due to processes such as ionization, while the second is more local and it accounts for the conversion between gas and stars.

We train all parameters as discussed in the previous section, using an L1 loss (see Eq.~\eqref{eq:loss}) as we find it produces the best results.

A full pipeline of the \jerald approach to produce HI maps, including the approximate N-body simulation and the steps to improve the DM distribution and to produce the stellar mass map, is illustrated in \figt \ref{fig:pipeline}. All free parameters in the model for DM, stellar mass and HI, are summarized in Table~\ref{tab:JERALDparams}.

\begin{table*}[t]
\begin{center}
\begin{tabular}{ c | c c c c c c }
\hline
\hline
Quantity & & Equation & Parameters & $N_p$ & $N_p^\text{tot}$ & $N_p^\text{used}$ \\
\hline
 & Displacement & \eqref{eq:displacement} & $\alpha_\text{dm}$, $\gamma_\text{dm}$ & $2 S_\text{dm}$ & \multirow{3}{*}{$5 S_\text{dm}$} & \multirow{3}{*}{$10$} \\
Dark Matter & Green's function & \eqref{eq:greenDM} & $\khDM$, $\klDM$, $\nDM$ & $3 S_\text{dm}$ \\
 & Transformation & - & - & - \\
 \hline
 & Displacement & \eqref{eq:displacement} & $\alpha_\text{sm}$, $\gamma_\text{sm}$ & $2 S_\text{sm}$ & \multirow{3}{*}{$5 S_\text{sm}+3$ ($+5 S_\text{dm}$)} & \multirow{3}{*}{13 ($+10$)} \\
Stellar mass & Green's function & \eqref{eq:greenSM} & $\khSM$, $\klSM$, $\nSM$ & $3 S_\text{sm}$ \\
 & Transformation & \eqref{eq:reluSM} & $w_\text{sm}$, $\mu_\text{sm}$, $b_\text{sm}$ & 3 \\
 \hline
\multirow{7}{*}{Neutral hydrogen} & \multirow{2}{*}{Displacement} & \multirow{2}{*}{\eqref{eq:displacementHI}} & $\alpha^{(1)}_\text{HI}$, $\gamma^{(1)}_\text{HI}$ & $2 S_\text{HI}$ & \multirow{7}{*}{\parbox{3cm}{$10 S_\text{HI}+11$\\ ($+5 S_\text{dm}+5 S_\text{sm}+3$)}} & \multirow{7}{*}{31 ($+10+13$)} \\
  &  &  & $\alpha^{(2)}_\text{HI}$, $\gamma^{(2)}_\text{HI}$ & $2 S_\text{HI}$ \\
 & \multirow{2}{*}{Green's function(s)} & \multirow{2}{*}{\eqref{eq:greenSM}} & $k^{h\,(1)}_\text{HI}$, $k^{l\,(1)}_\text{HI}$, $\nu^{(2)}_\text{HI}$ & $3 S_\text{HI}$ \\
&  &  & $k^{h\,(2)}_\text{HI}$, $k^{l\,(2)}_\text{HI}$, $\nu^{(2)}_\text{HI}$ & $3 S_\text{HI}$ \\
 & \multirow{3}{*}{Transformation} & \eqref{eq:reluHI} & $w_\text{HI}$, $\mu_\text{HI}$, $b_\text{HI}$ & 3 \\
 &  & \eqref{eq:depletion} & $\beta^{(1)}$, $\beta^{(1)}$, $\eta$, $\gamma^{(3)}_\text{HI}$ & 4 \\
  &  & \eqref{eq:dfilter} & $k^h_\text{I}$, $k^l_\text{I}$, $\nu_\text{I}$, $\xi$ & 4 \\
  \hline
\end{tabular}
\caption{\jerald parameters for DM, stellar mass and HI targets, at a given redshift. $N_p$ is the number of parameters in each part of the model and $N_p^\text{tot}$ the total number of parameters for each predicted quantity, with $S_\text{dm}$, $S_\text{sm}$ and $S_\text{HI}$ being the number of displacement layers used for DM, stellar mass and HI respectively (all set to 2 except for $S_\text{sm}=3$ at $z=5$). $N_p^\text{used}$ is the total the number of parameters used in this work (except for stellar mass at $z=5$, where $N_p^\text{used}=18\ (+10)$). For stellar mass and HI, the total number of parameters also includes, in brackets, those used to improve the DM distribution (for both) and to produce the stellar mass map (for HI).}
\label{tab:JERALDparams}
\end{center}
\end{table*}

\section{Code specifications and availability}
\label{sec:code}

\begin{figure}[ht]
    \centering
    \includegraphics[width=0.95\linewidth]{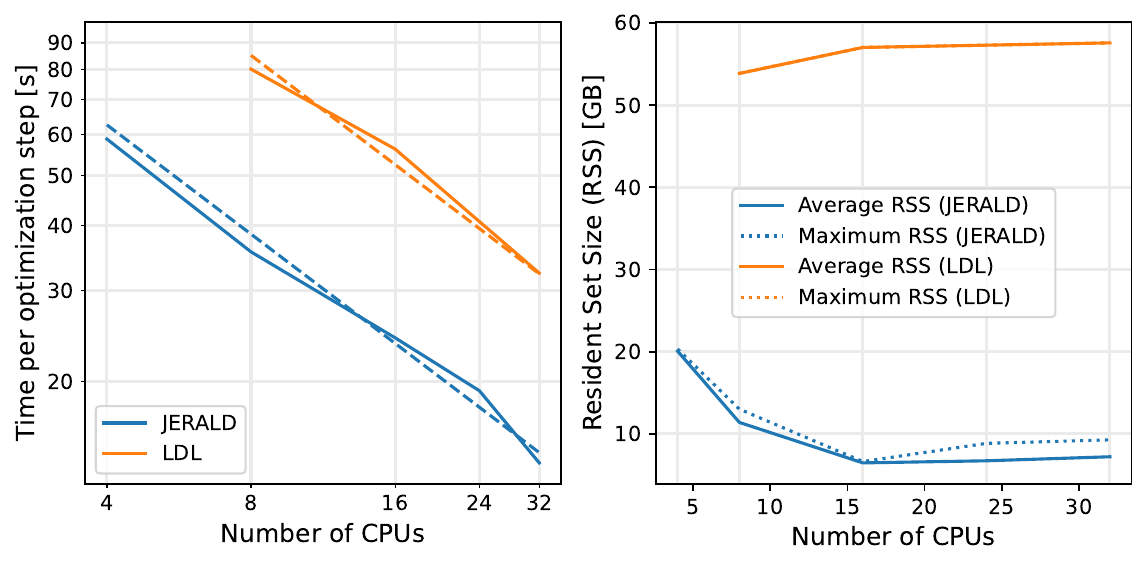}
    \caption{Time per optimization step (left) and resident set size (RSS, right, a measure of memory used in GB) for \jerald and the original LDL code from~\cite{LDL} on one node of the Leonardo supercomputer. The RSS is given by the slurm tool \texttt{sacct}, which may not capture rapid memory usage spikes. Times are given as averages over 20 optimization steps in the case of $512^3$ particles and a $512^3$ voxels DM target map and 1 displacement layer, and the RSS values are those associated with the 20 steps. The dashed lines in the plot on the left indicate the power law $\propto (N_\text{CPUs})^{-0.7}$ for reference. The maximum RSS for LDL is not visible as it coincides with the average RSS.}
    \label{fig:comparison}
\end{figure}

The \jerald code is written partly in Python using \texttt{JAX} \cite{jax} and partly in \texttt{C++}. The CIC paint and readout---i.e. the algorithms to compute density fields from properties defined at particle positions and vice-versa---and the DFTs are implemented in \texttt{C++}, the latter via MPI-enabled FFTW \cite{fftw}, and bound to Python via Pybind11 \cite{pybind11}. In Python, CIC paint and readout as well as forward and backward DFTs are defined as JAX primitives and equipped with definitions of VJPs, making them jittable and available for automatic differentiation. All methods are parallelized using MPI (in Python, via MPI4JAX \cite{mpi4jax}). Each MPI rank deals with approximately $1/W$ of the particles and $1/W$ of the density fields, where $W$ is the size of the MPI communicator, making the code scale approximately inversely proportional in time and constant in memory as $W$ increases (see Figure \ref{fig:comparison}). For the $512^3$ particles and $512^3$ mesh points maps discussed in this work, each step in the optimization (evaluation of training and validation loss and training loss gradient) takes $\sim$15$\div$20s per displacement layer in one node of the Leonardo supercomputer in CINECA (32 CPUs). The number of optimization steps depends on the kind of reference map and on the optimizer parameters; generally, at most 500 steps are sufficient for the validation loss to reach convergence, and one training typically takes $\sim$2$\div$5hrs, depending on the number of displacement layers (in this work, we used at most 3). Evaluation of the maps with a pre-trained model takes $\sim$10s with the same resources. 

For a comparison of speed and memory usage with the original LDL code from~\cite{LDL}, in Figure \ref{fig:comparison} we show time per optimization step (left) and resident set size (right) for 20 training steps in the case of a $512^3$ voxels DM target with $512^3$ particles and one displacement layer (the number of CPUs is also the size of the MPI communicator $W$). We observe that, while the scaling law for the time per optimization step is approximately the same, the proportionality coefficient for \jerald is significantly smaller with respect to the one of LDL; nonetheless, the curves for the RSS indicate that \jerald is also more memory efficient than LDL.

The \jerald code, together with installation requirements and instructions, is available at \url{https://github.com/maurorigo/JERALD/tree/main}. 

\section{Results}
\label{sec:results}

We now turn to evaluating \jerald's performance. We first give details about the reference simulations and the training procedure, then we investigate the quality of \jerald's outputs against the same maps used for training, and finally we provide a first assessment of its generalization capabilities by applying the trained model to an approximate $N$-body simulation with new initial conditions (ICs) and compare the output maps with targets that were not seen during training. 

\subsection{Simulations and training}\label{sec:sims}

The reference maps used in the subsequent section were extracted from the Sherwood-Relics simulation suite \cite{Sherwood,sherwood-relics}. This set of simulations follows the evolution of DM, gas and stars in a periodic cosmological volume, and the key physical ingredients are: $i)$ an evolving UV background that returns a thermal history for the IGM which is in agreement with observations; $ii)$ a simplified star formation criterion which converts into star particles all the gas particles whose overdensities are larger than 1000 times the mean baryon density and whose temperatures are below 10$^5$ K.
These simulations are supposed to offer an accurate description of the filamentary cosmic web, and in particular \lya forest statistics in agreement with observations \cite{irsic24}. The specific simulations used here feature a box size of 80$h^{-1}$Mpc and $1024^3$ DM and gas particles, with fiducial parameters $\Omega_m = 0.308$, $\Omega_\Lambda = 0.692$, $\Omega_b = 0.0482$, $h = 0.678$, $\sigma_8 = 0.829$, and $n_{\rm s} = 0.961$.

Input DM particles positions for \jerald are computed using the approximate N-body solver FastPM \cite{fastpm}; we generate the ICs on a $(2\times512)^3$ mesh, then evolve $512^3$ particles on a $512^3$ mesh using 2LPT up to $z=9$, and finally run 10 FastPM steps up to the desired redshift values ($z=0,2,5$). Starting from the FastPM input, \jerald's outputs were computed using the same $512^3$ particles and a $512^3$ mesh for all fields, making the resolution of the output maps 8$\times$ smaller than the reference simulation. 

We use the Adam optimizer \cite{adam} with a scheduled learning rate ranging from 0.002 to 0.02 after extensive experimentation with other optimizers and hyperparameters. Model parameters are initialized to arbitrary values, and, in the code, they are multiplied by constants to loosely guarantee them to vary in the same range for a more stable training procedure. Specifically, all parameters are expected to be $\mathcal{O}(1)$ except for the coefficients in front of the potential terms (e.g. $\alpha$ in Eq.~\eqref{eq:displacement}), which are generally of order $10^{-2}$. Yet, parameters that are initially considerably far from the optimal ones may be associated with very large gradients; for this reason, we adopt a time-varying schedule for the learning rate $\eta(i)$, increasing it during training following a softmax schedule:
\begin{equation}
    \eta(t) = (\eta(N_\text{step})-\eta(0))\left(\frac{2}{e^{-t/\tau_\eta}+1}-1\right) + \eta(0)
\end{equation}
where generally, for $N_\text{step}$ training steps, $\eta(0)\sim\mathcal{O}(10^{-3})$, $\eta(N_\text{step})\sim\mathcal{O}(10^{-2})$ and $\tau_\eta\sim100$.

The training/validation/test sets splitting is chosen as follows: for a $M\times M\times M$ map, where $M$ is the number of voxels in each direction, we use a $(0.5 M)\times M\times M$ box for training, a $(0.43 M)\times(0.43 M)\times(0.43 M)$ box for testing, and the rest for validation, which correspond to approximately 50\% of the pixels in the original map for training, 42\% for validation and 8\% for testing. 

As shown in Table \ref{tab:JERALDparams}, all results included in the next sections were obtained using 2 displacement layers for each target at each redshift, except in the case of stellar mass for $z=5$, where 3 displacement layers were utilized. We adopted a unit $n$ hyperparameter for the smoothing operator in Eq.~\eqref{eq:loss2} in the case of stellar mass at $z=0$ and $z=2$ and HI at $z=5$, while for stellar mass at $z=5$ $n=0.7$ was found to give the best result. In order to bring all training values to the same scale, of order unity, all calculations involving stellar mass were performed in units of $10^{10}h^{-1}M_\odot$, while target HI number densities were divided by a redshift-dependent constant. 

\subsection{Evaluation of \jerald's output maps}

\begin{figure*}[ht]
    \centering
    \includegraphics[width=1.\textwidth]{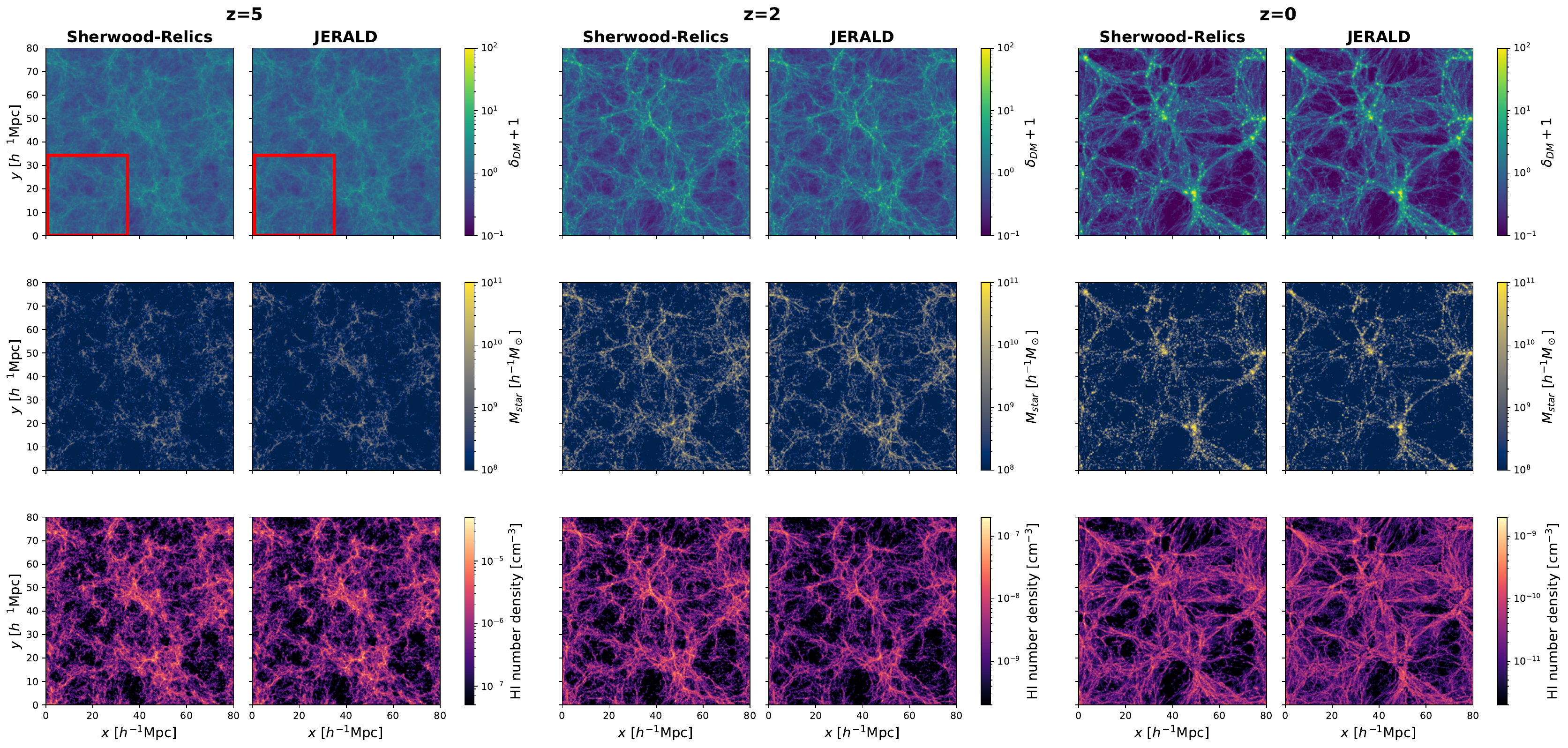}
    \caption{Dark matter overdensity, stellar mass and neutral hydrogen number density maps, from top to bottom. Each pair of maps compares the reference map from Sherwood-Relics (left panel) with \jerald's output (right panel), for $z=5$, $z=2$ and $z=0$ (left to right). Notice the different range in the colormaps for neutral hydrogen. All images were obtained as 10$h^{-1}$Mpc $z$-slices of the full $(80h^{-1}\text{Mpc})^3$ simulation box. All voxels used to produce the images were part of either the validation or the test set; the red rectangles in the top left pair of maps---omitted in the others---represent the test sub-box, and it extends for the full length of the $z$-slice used to produce these images. Note that the HI map for $z=0$ was normalized with the average of the reference map.}
    \label{fig:maps}
\end{figure*}

Figure \ref{fig:maps} shows slices of dark matter overdensity, stellar mass and neutral hydrogen number density maps obtained using \jerald from an input approximate DM distribution from FastPM at redshift $z=0$, $z=2$ and $z=5$, compared with the respective Sherwood-Relics references. The red square shows the test region -- i.e., the part of the cube that has not been used for training nor validation. All the voxels used to produce the images in the figure were either part of the validation or the test sets. The visual agreement is remarkable.

\begin{figure}[ht]
    \centering
    \includegraphics[width=.48\textwidth]{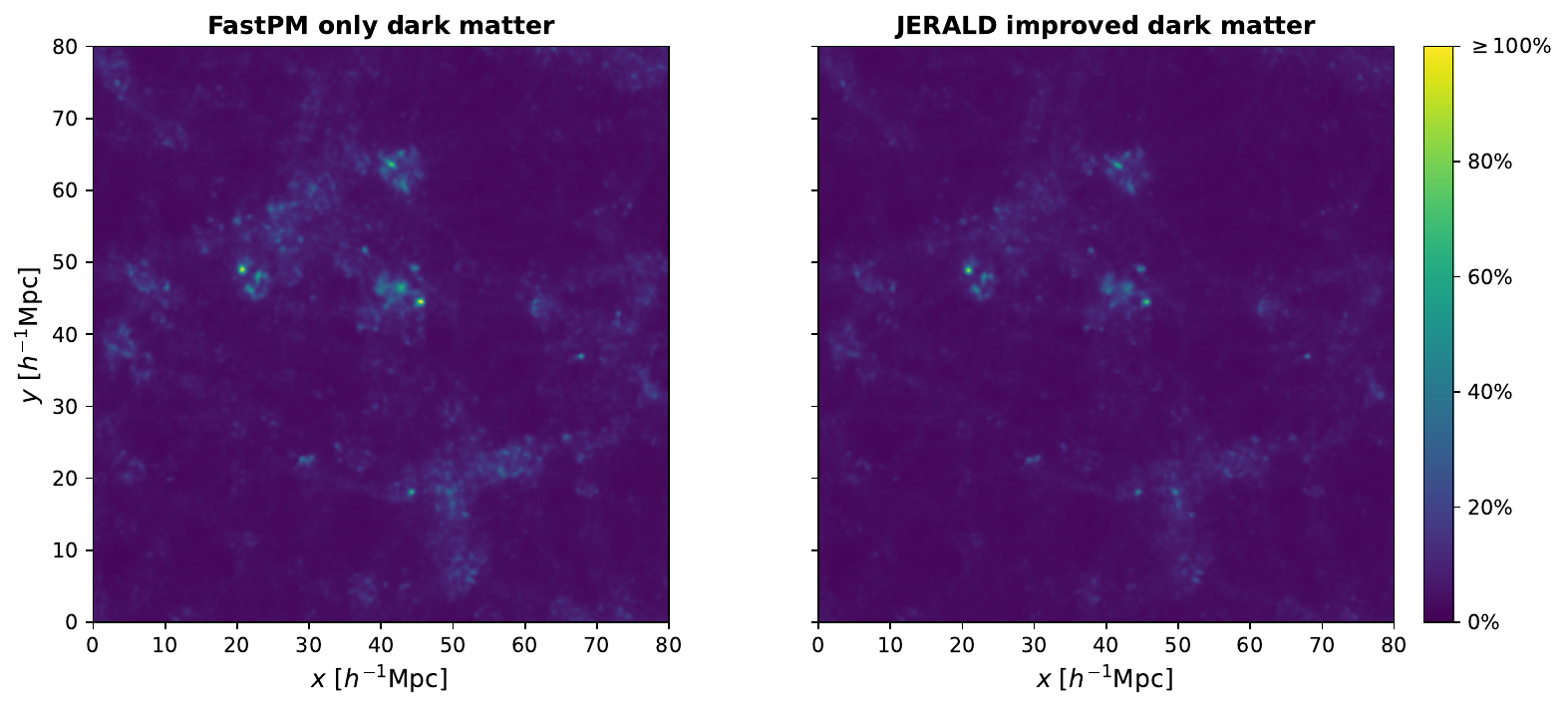}
    \includegraphics[width=.48\textwidth]{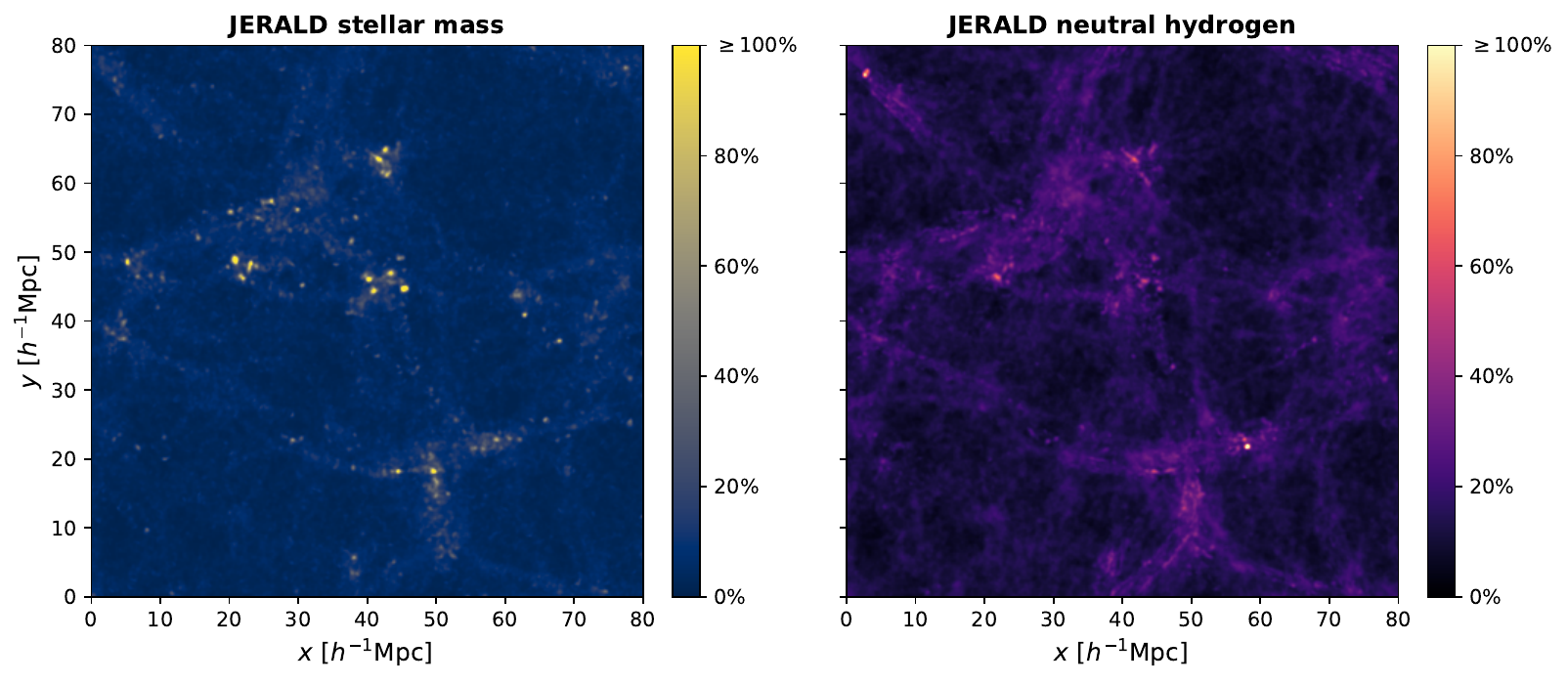}
    \caption{Relative difference between $z=0$ FastPM only and \jerald's output for dark matter, stellar and neutral hydrogen maps. Each map used to compute the relative differences was clipped below 0.5 times the average mass of the reference map and smoothed with a Gaussian filter with a radius of $200h^{-1}$kpc. The plots show the relative difference averaged over the entire length of the $80h^{-1}$Mpc box along the $z$ direction.}
    \label{fig:reldiffplots1}
\end{figure}

\begin{figure}[ht]
    \centering
    \includegraphics[width=.48\textwidth]{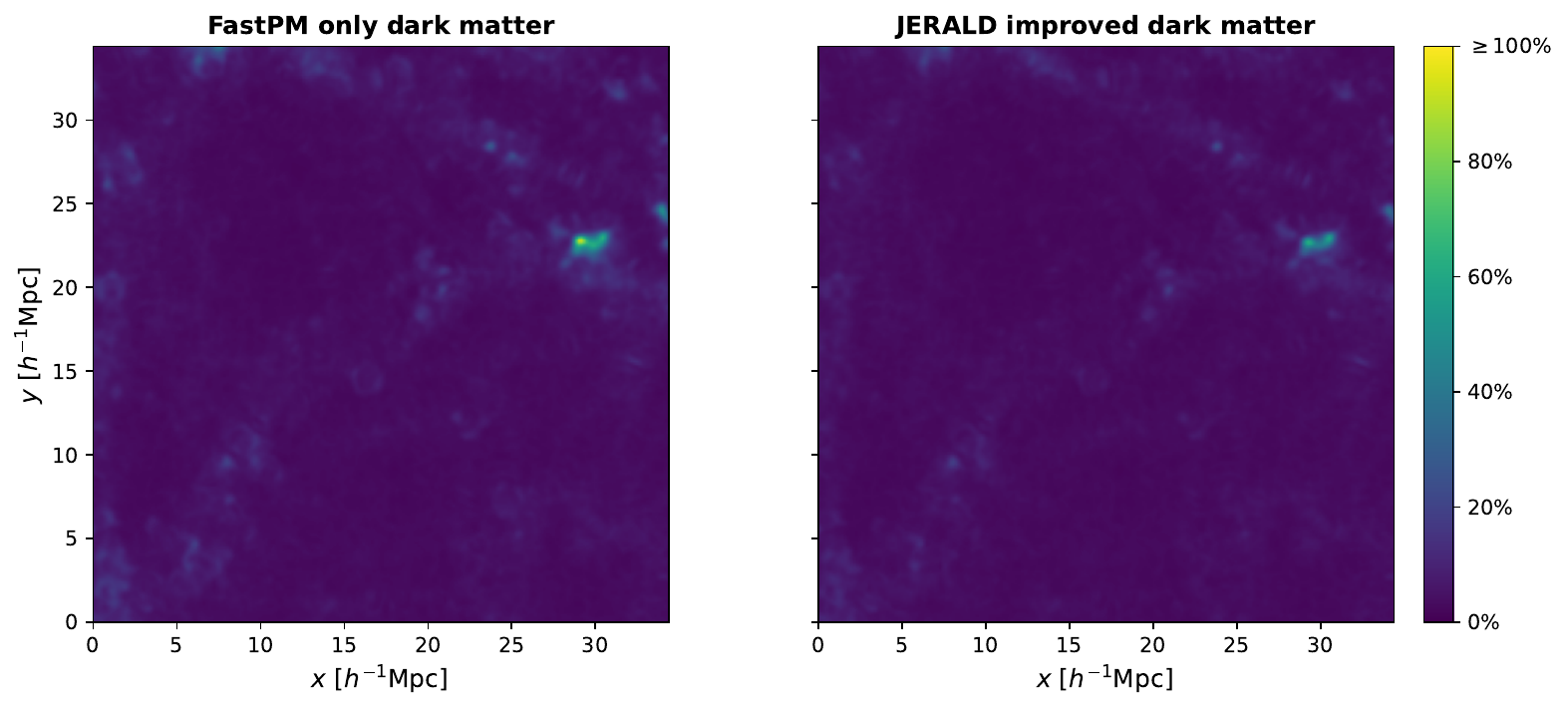}
    \includegraphics[width=.48\textwidth]{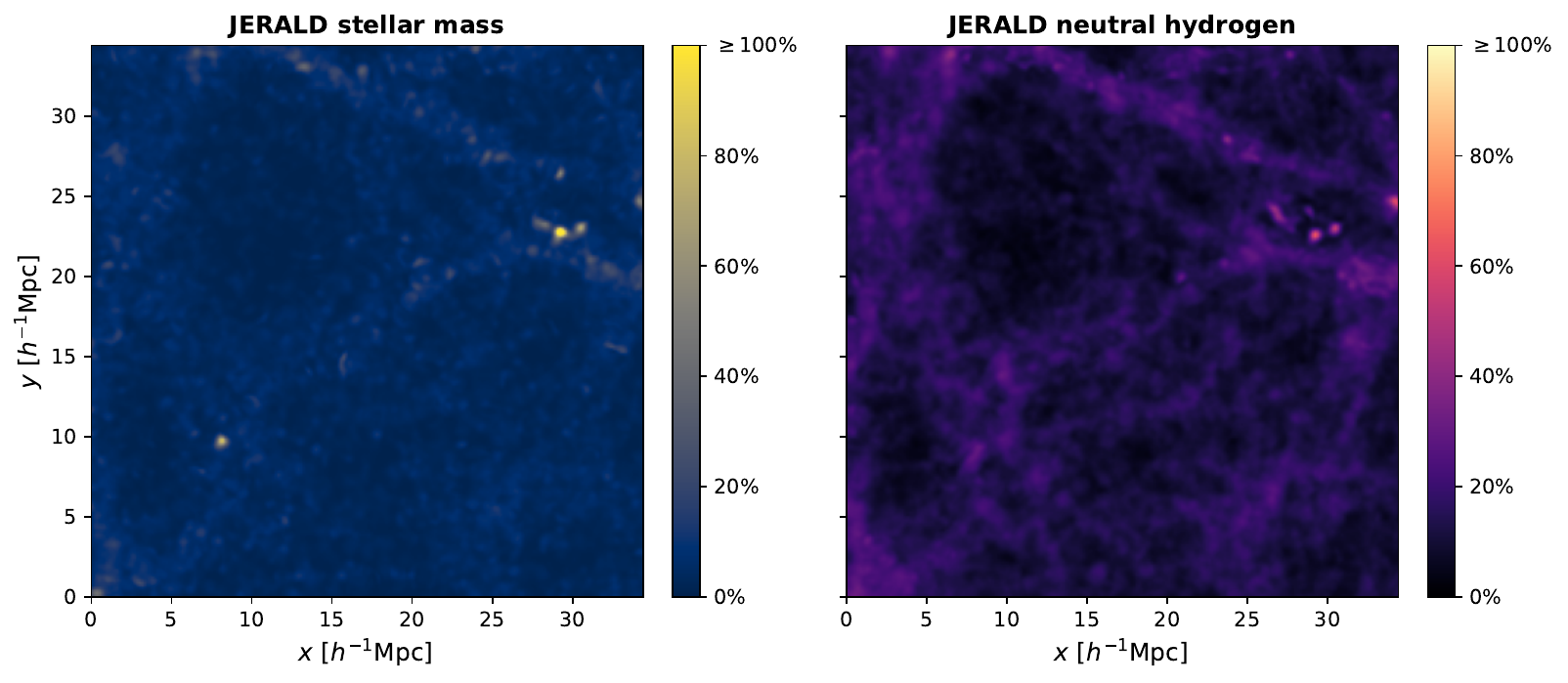}
    \caption{As in Fig.~\ref{fig:reldiffplots1}, but showing the relative difference only for the test set, averaged over the $34.4h^{-1}$Mpc of the test sub-box along the $z$ direction.}
    \label{fig:reldiffplots2}
\end{figure}

To evaluate more quantitatively the performances of \jerald, in \figts \ref{fig:reldiffplots1} and \ref{fig:reldiffplots2} we also show the relative difference between the reference maps and \jerald's outputs, both post-processed in two steps. Firstly, each map was clipped below $0.5\, \bar \phi_i$, where $\bar \phi_i,\,i=\text{dm/sm/HI}$ are the averages of the reference Sherwood-Relics DM, stellar mass or HI maps. This is because we expect mesh points with small density/mass to only bring minor contributions to the quantities we plan to compute using maps produced by \jerald. By construction, the loss in Eq.~\eqref{eq:loss} prioritizes matching mesh points with large values; for this reason, mesh points with small values may produce large relative differences, which could give a misleading impression of poor performance. The clipping threshold is chosen arbitrarily and it relies on the knowledge of the averages of the reference maps $\bar \phi_i,\,i=\text{dm/sm/HI}$ only to define a value that is small \textit{relative to each map type}; of course, when using \jerald on unseen maps, these averages would not be available nor necessary. Secondly, each map was smoothed using a Gaussian filter with a radius of $200h^{-1}$kpc. We only illustrate the relative differences for $z=0$, as distributions become progressively more nonlinear for lower redshifts and therefore more complicated to model---the relative difference maps for $z=2$ and $z=5$ display considerably less features than the ones illustrated here. Despite the high resolution and small smoothing radius, the relative differences are characterized by only several notable features, with most of the maps lying below 20\%. Note that the locations of the prominent features present in the maps are approximately the same.

\setlength{\tabcolsep}{5pt}
\renewcommand{\arraystretch}{1.2}

\begin{figure}[ht]
    \centering
    \includegraphics[width=0.45\textwidth]{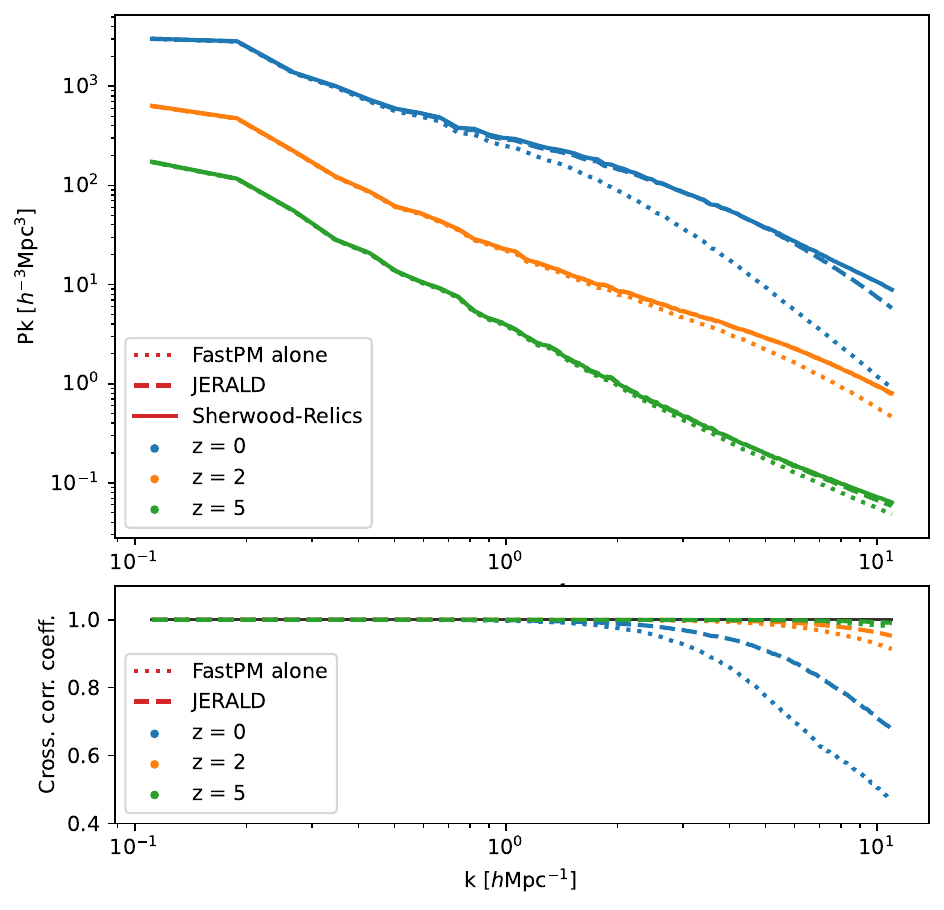}
    \caption{DM power spectra (top) and cross correlation coefficients (bottom) for the reference Sherwood-Relics simulation (solid), the approximate N-body FastPM simulation (dotted) and the map improved by \jerald (dashed), for three different redshifts. For $z=5$, the improved power spectrum is almost not visible as it overlaps with the reference one.}
    \label{fig:DMplots}
\end{figure}

\begin{figure}[ht]
    \centering
    \includegraphics[width=0.45\textwidth]{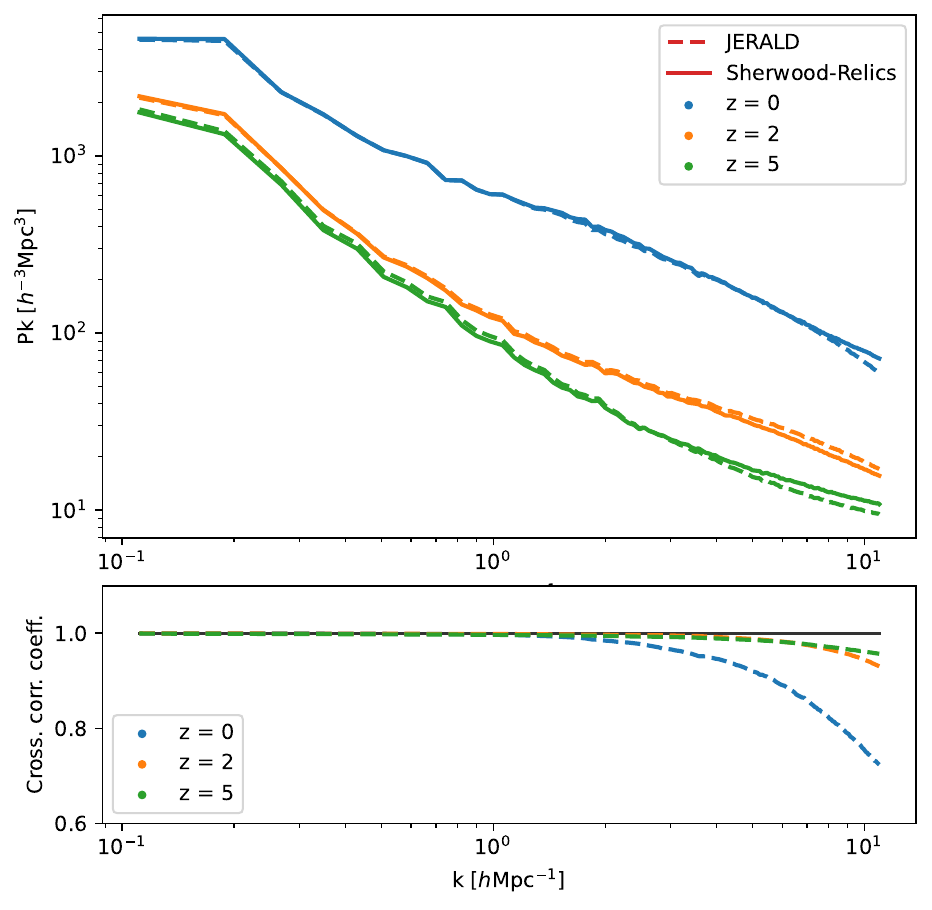}
    \caption{Stellar power spectra (top) and cross correlation coefficients (bottom) for the reference Sherwood-Relics simulation (solid) and the map produced using \jerald (dashed), at three different redshifts.}
    \label{fig:Mstarplots}
\end{figure}

\begin{figure}[ht]
    \centering
    \includegraphics[width=0.45\textwidth]{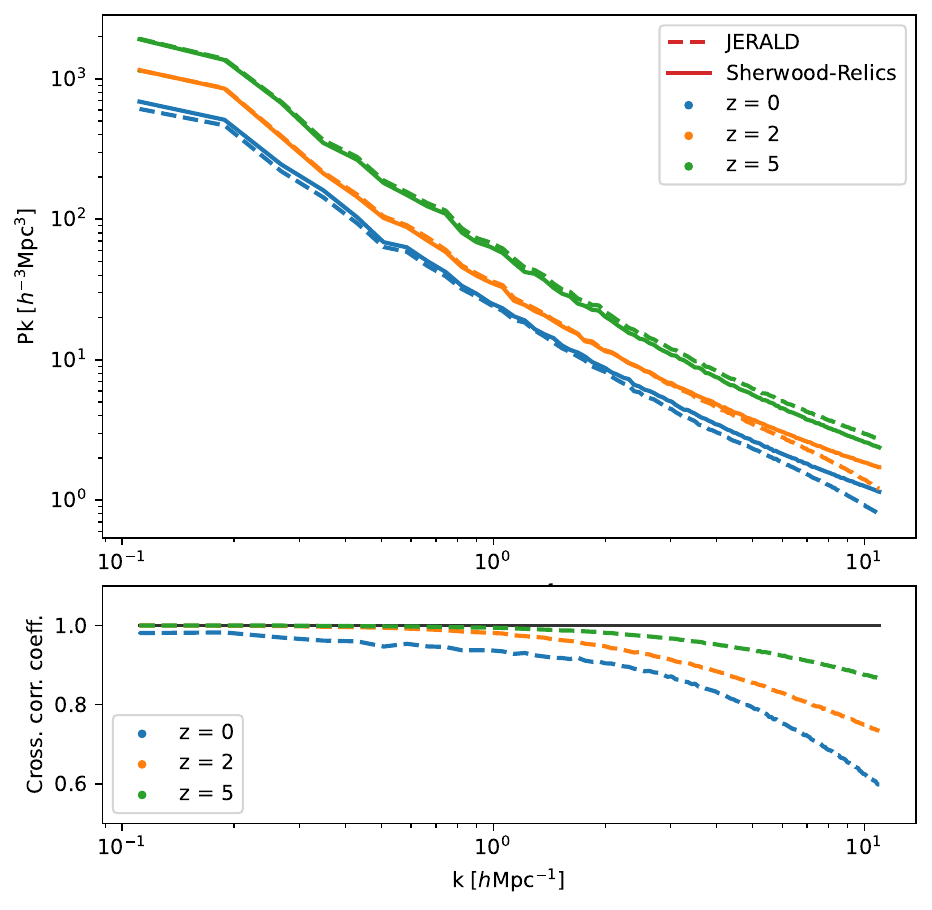}
    \caption{Low-intermediate density neutral hydrogen power spectra (top) and cross correlation coefficients (bottom) for the reference Sherwood-Relics simulation (solid) and the map produced using \jerald (dashed), at three different redshifts. For visualization purposes, the green curves ($z=5$) in the top panel were offset by multiplying them by an arbitrary constant (1.7), otherwise they would overlap with the orange ones ($z=2$). The correct scale of the power spectra at $z=5$ is the same as for $z=2$.}
    \label{fig:nHIplots}
\end{figure}

To analyze the statistics of \jerald's outputs, we also compare the power spectra of the maps with the reference ones in \figts \ref{fig:DMplots}, \ref{fig:Mstarplots} and \ref{fig:nHIplots} for dark matter, stellar and neutral hydrogen overdensities respectively. Similarly to \cite{LDL}, the power spectra were computed on the entire (80$h^{-1}$Mpc)$^3$ simulation box, rather than on the test subset only, in order to capture the large scales as well and determine whether the model affects the distributions at those scales. Including the training data in this analysis may bias the results, therefore in the next section we will also analyze the performance of the approach on unseen maps. The bottom panel of \figts\ \ref{fig:DMplots}, \ref{fig:Mstarplots} and \ref{fig:nHIplots} shows the cross correlation coefficient, calculated as:
\begin{equation}
    c(k) = \frac{P_\text{cross}(k)}{\sqrt{P_\text{ref}(k)P_\text{\jerald}(k)}}
\end{equation}
where $P_\text{cross}(k)$ is the cross power spectrum and $P_\text{ref}(k)$ and $P_\text{\jerald}(k)$ are the power spectra of the Sherwood-Relics reference simulation and \jerald, respectively. In each case and at each redshift, the power spectra of the maps produced by \jerald are in excellent agreement with the reference ones even up to highly non-linear scales. As expected, the cross correlation coefficients show that the agreement between the maps increases as redshift increases. Note that, in computing the stellar overdensities, we used the average of the reference maps rather than that of the maps produced by \jerald. This was done because, when modeling stellar mass, the introduction of the smoothing filter of Eq.~\eqref{eq:loss2} in the loss leads to sharp mass peaks being smeared out and, as a consequence, the average of the map produced by \jerald is smaller than the one of the reference (though only by a small factor of $\sim5\%$). Nonetheless, these mass peaks do not bring significant contributions to the overall shape of the power spectra, as \figt\ \ref{fig:Mstarplots} highlights, and we expect them not to pose problems in applications of this method.

\subsection{Generalization capabilities}
\label{sec:generalization}

Until now, we presented results evaluated on the same reference maps used in the training procedure. Despite the fact that only 50\% of the grid points in each map are used for training and the model has only few parameters, the power spectra and relative difference plots reported above may still be affected by overfitting; for this reason, in this section we investigate the performance of \jerald on unseen simulations. Specifically, we run a new full-hydrodynamic simulation using the same code as the one used for the reference simulation of the previous sections, same box and mesh sizes and same cosmological and astrophysical parameters, but a different seed for the generation of the ICs. We then run FastPM as in Sec. \ref{sec:sims} with the new ICs and apply \jerald with fixed parameters, trained on the simulation used in the previous sections. 

\begin{figure*}[htb]
    \centering
    \includegraphics[width=1.\textwidth]{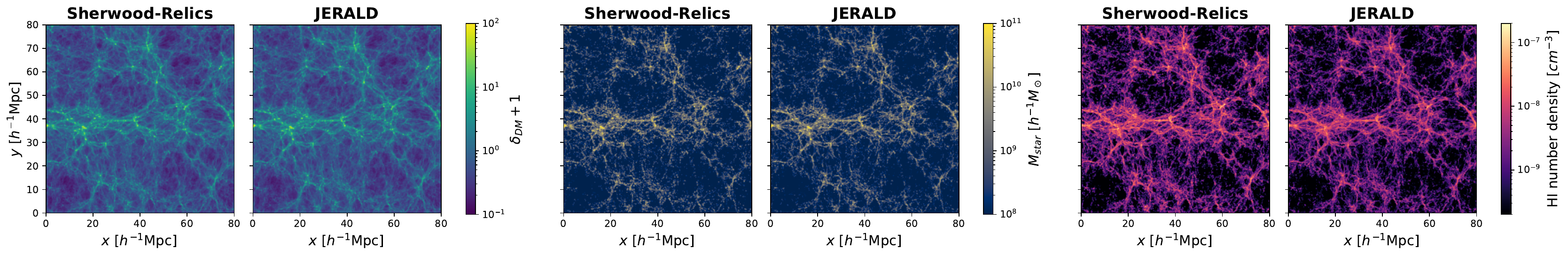}
    \caption{Similar to \figt \ref{fig:maps}, but for a new Sherwood-Relics simulation unseen by \jerald. In this case, all plots correspond to $z=2$ and the different quantities---i.e., DM overdensity, stellar mass and HI number density---are arranged horizontally. In this case, no test sub-box is highlighted, because \jerald was pre-trained on another simulation.}
    \label{fig:mapsseed}
\end{figure*}

\begin{figure*}[ht]
    \centering
    \includegraphics[width=1.\textwidth]{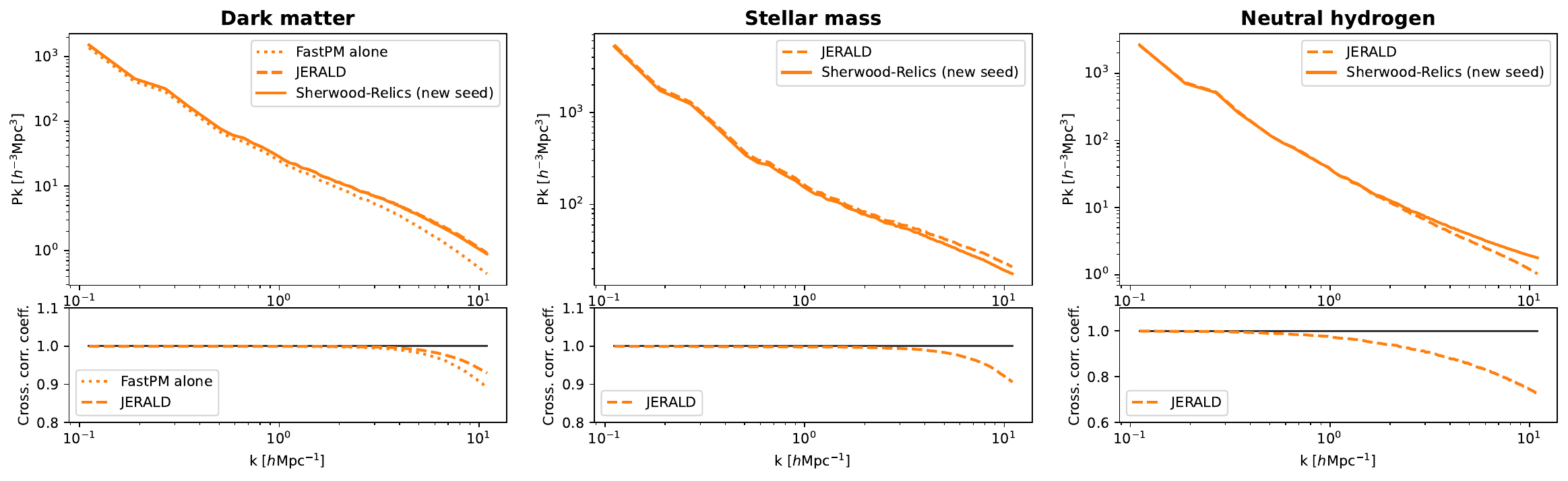}
    \caption{Dark matter, stellar and neutral hydrogen power spectra at $z=2$ for a simulation with box size (80$h^{-1}$Mpc)$^3$ and mesh size $1024^3$ with different initial conditions with respect to the one on which \jerald's parameters were trained.}
    \label{fig:pkseed}
\end{figure*}

\figt \ref{fig:mapsseed} shows dark matter overdensity, stellar mass and neutral hydrogen number density maps as in \figt \ref{fig:maps} with the new unseen simulation limited to just $z=2$ for computational reasons. The visual agreement is very good also in this case. In \figt \ref{fig:pkseed} we show the associated power spectra and cross correlation coefficients. We compare this new reference (solid) with \jerald's output (dashed), and we observe similar performance as what was obtained on the reference simulation on which \jerald was trained. This is a strong indication of the method's generalization capabilities, which we plan to investigate in more detail in future works, when we will also evaluate the performance for previously unseen simulations with different astrophysical and cosmological parameters as well as transferability of results to different simulation codes. 

\section{Conclusions}\label{sec:conclusions}

We presented a new, improved implementation of a physically motivated approach to reducing the enormous computational cost of cosmological simulations in order to obtain accurate dark matter and baryonic maps. Our code, \jerald (JAX Enhanced Resolution Approximate Lagrangian Dynamics), builds on the Lagrangian Deep Learning method originally proposed in Ref.~\cite{LDL}, and focuses in particular on producing, on top of more accurate DM and stellar maps, high-fidelity neutral hydrogen maps. 

We showed that our improved implementation is capable of producing maps in excellent agreement, in terms of power spectra for DM, stars and HI, with full-hydrodynamical simulations, up to several $h$Mpc$^{-1}$. The performances of \jerald are more impressive at higher redshift ($z\geq2$), where non-linearities are less prominent: in this regime, the agreement, in terms of power spectra, is above the 70\% level all the way to very small scales ($k\sim 20\,h$Mpc$^{-1}$). This opens the door to investigating, in quantitative detail, the impact on very small scales of different dark matter models. 

\jerald was developed from scratch, with the aim of being user-friendly, fully parallel and highly computationally efficient. Its ultimate goal is not limited to producing baryonic maps: we aim at exploiting this approach to study observable HI-related quantities such as the \lya forest and the 21cm intensity \cite{villa18}. These two observables are particularly important, as neutral hydrogen is an excellent probe of DM models beyond cold dark matter both in emission and absorption \cite{carucci15,irsic24}---an alternative that could in principle alleviate some of the small-scale inconsistencies of the \lcdm model: for example, a lighter (``warm'') DM particle could offer the solution to the core/cusp problem, as the main effect of its larger velocity is to ``smear out" structures at small scales \cite{wdm,wdmreview}. The quantitative study of such models from large scale structure and \lya forest, as well as 21cm mapping, can benefit from being embedded in a neural network-enabled simulation-based inference (SBI) framework (e.g., \cite{Alsing2019,Makinen2022TheCG,Makinen2024HybridSS,Charnock2018AutomaticPI}), full high-dimensional likelihood evaluation \citep{osti_1523502,Dai2022TranslationAR} or field-level inference~(e.g.~\cite{Porqueres2021}) approaches, which have been shown to extract a great deal more information from the data than traditional hand-crafted summaries such as the power spectrum. Often, the bottleneck to the full exploitation of such techniques lies in the limited amount of high-resolution simulations on which to train the neural models. With \jerald, we have introduced a tool that will enable us to take a major step in future work to enable such studies. 


\section*{Acknowledgements}
RT acknowledges co-funding from Next Generation EU, in the context of the National Recovery and Resilience Plan, Investment PE1 – Project FAIR ``Future Artificial Intelligence Research''. This resource was co-financed by the Next Generation EU [DM 1555 del 11.10.22]. RT is partially supported by the Fondazione ICSC, Spoke 3 ``Astrophysics and Cosmos Observations'', Piano Nazionale di Ripresa e Resilienza Project ID CN00000013 ``Italian Research Center on High-Performance Computing, Big Data and Quantum Computing'' funded by MUR Missione 4 Componente 2 Investimento 1.4: Potenziamento strutture di ricerca e creazione di ``campioni nazionali di R\&S (M4C2-19)'' - Next Generation EU (NGEU). The Sherwood and Sherwood-Relics simulations used in this work were possible thanks to the Partnership for Advanced Computing in Europe (PRACE), during the 16th Call. The Sherwood-Relics simulations were also performed using time allocated during the Science and Technology Facilities Council (STFC) DiRAC 12th Call. We used the Cambridge Service for Data Driven Discovery (CSD3), part of which is operated by the University of Cambridge Research Computing on behalf of the STFC DiRAC HPC Facility (www.dirac.ac.uk). The DiRAC component of CSD3 was funded by BEIS capital funding via STFC capital grants ST/P002307/1 and ST/R002452/1 and STFC operations grant ST/R00689X/1. Part of the simulations were postprocessed on Ulysses supercomputer at SISSA.
MV and RT are also partially supported by INFN INDARK grant.

\bibliography{biblio}

\end{document}